\documentclass{emulateapj}

\usepackage{amsbsy}

\def\gsim{\;\rlap{\lower 2.5pt
\hbox{$\sim$}}\raise 1.5pt\hbox{$>$}\;}
\def\lsim{\;\rlap{\lower 2.5pt
\hbox{$\sim$}}\raise 1.5pt\hbox{$<$}\;}

\def\etal{{et al.\thinspace}}
\def\eg{{\it e.g.\ }}

\newcommand{\be}{\begin{equation}}
\newcommand{\ba}{\begin{eqnarray}}
\newcommand{\ee}{\end{equation}}
\newcommand{\ea}{\end{eqnarray}}

\begin{document}

\title{Understanding Galaxy Outflows as the Product of Unstable Turbulent Support}

\author{Evan Scannapieco}
\affil{School of Earth and Space Exploration,  Arizona
  State University, P.O.  Box 871404, Tempe, AZ, 85287-1404.}
  
\begin{abstract}

The interstellar medium is a multiphase gas in which turbulent support is as important as thermal pressure.  Sustaining this configuration requires both continuous turbulent stirring  and continuous radiative cooling to match the decay  of turbulent energy.  While this equilibrium can persist for small turbulent  velocities, if the one-dimensional velocity dispersion is larger than $\approx 35$ km/s, the gas moves into an unstable regime that leads to rapid heating.  I study the implications of this turbulent runaway, showing that it causes a hot gas outflow to form in all galaxies with a gas surface density above  $\approx 50$ M$_\odot$ pc$^{-2},$ corresponding to a star formation rate per unit area of $\approx 0.1$ M$_\odot$ yr$^{-1}$ kpc$^{-2}.$  For galaxies with $v_{\rm esc} \gsim 200$ km/s, the sonic point of this hot outflow should lie interior to the region containing cold gas and stars, while for galaxies with smaller escape velocities, the sonic point should lie outside this region.  This leads to efficient cold cloud acceleration in higher mass galaxies, while in lower mass galaxies, clouds may be ejected by random turbulent motions rather than accelerated by the wind.   Finally,  I show that energy balance cannot be achieved at all for turbulent media above a surface density of $\approx 10^5$ M$_\odot$ pc$^{-2}.$ 

\end{abstract}


\section{Introduction}

Galaxy outflows are well observed over many masses and redshifts (\eg Martin 1999; Pettini \etal 2001; Veilleux, Cecil  \& Bland-Hawthorn 2005) and  form in all galaxies in which the star formation rate per unit area, $\dot \Sigma_*,$ exceeds the ``Heckman limit" of $\dot \Sigma_\star^{\rm crit} \approx 0.1 M_\odot$ yr$^{-1}$ kpc$^{-2}$ (Heckman 2002). They shape the galaxy mass-metallicity relation (\eg Tremonti et al. 2004), affect the gas content and number density of dwarf galaxies (\eg Scannapieco \etal 2001; Benson \etal 2003), and enrich the intergalactic medium (\eg Scannapieco \etal 2006).  Yet, they are extremely difficult to model theoretically, as a complete model of galaxy outflows requires  a deep understanding of both star formation and the evolution of the multiphase interstellar medium (ISM). 

Despite these complexities, I show in this {\em Letter} that   many of the features of galaxy outflows can be understood from a more general perspective.   Regardless of how stars impact their environment, a generic result of their evolution is the turbulent stirring of the ISM, and the resulting turbulent support is at least as important as thermal or magnetic pressure (Elmegreen \& Scalo 2004).  Unlike other sources of support however, turbulence requires both continuous energy input, as it decays away to thermal energy, and continuous radiative cooling, as this thermal energy must be removed from the system.   Here I show that this equilibrium is stable only when $\dot \Sigma_\star \lesssim 0.1 M_\odot$ yr$^{-1}$ kpc$^{-2},$ while galaxies with larger $\dot \Sigma_\star$ values will develop  outflows with many of the key features observed in optical and X-ray studies. 

The structure of this work is as follows.  In \S2, I show how $\dot \Sigma_\star^{\rm crit}$  arises from a turbulent heating instability that occurs due to the decreasing efficiency of radiative cooling above $10^{5.5}$K.  In \S3, I explore the implications of this instability  for the properties of both starbursting galaxies and the hot and cold phases observed in galaxy outflows (where ``cold" means $\leq 10^4$K in this context). Finally, in \S4, I describe an extreme case in which turbulent heating cannot be sustained for a single dynamical time, and relate it to the maximum observed stellar density.
 
 \section{Critical Surface Density}

In this section, I show that  many of the features of outflowing galaxies can be understood as arising from  three underlying  properties:  hydrostatic equilibrium,  turbulent support, and self-regulation.  Each of these concepts can be formalized quantitatively. 

Hydrostatic equilibrium relates  pressure and  density as
$\frac{dp}{dz} \approx \frac{p}{H} \approx 4 \pi \rho_{\rm g}  G \Sigma_0,$
where $p$ and $\rho_g$ are the pressure and gas density in the midplane of a disk (or center of a spherical distribution), $H$ is the gas scale height, 
$\Sigma_0$ is the total matter surface density, and $G$ is the gravitational constant.   
This can be rewritten as a function of the gas surface density, $\Sigma_g$ as 
\be
p \approx 4 \pi G \Sigma_g  \Sigma_0.
\label{eq:he}
\ee

Turbulent support can be formalized by requiring that turbulence provides a fixed and substantial fraction $f$ of the total pressure:
\be
f p = \rho_g \sigma_{1D}^2,
\label{eq:pturb}
\ee
where $\sigma_{1D}$ is the average 1D turbulent velocity dispersion.   In a galaxy, $\sigma_{1D}$ can often be supersonic (Mac Low \& Klessen 2004), such that $f \approx 1,$  but it cannot be extremely subsonic, such that $f \ll 1$ (\eg Elmegreen \& Scalo 2004; Dalcanton \& Stilp 2010).

Finally, in a disk galaxy one can formalize the concept of self-regulation with the Toomre parameter, $Q,$ which is above 1 if
the  pressure is sufficient to stabilize perturbations on scales too small to be stabilized by rotation.  In a pure gas disk,
 $Q_{\rm gas}  =  {\sigma_{1D} \kappa}(f^{1/2} \pi G \Sigma_g)^{-1},$  where $\kappa$ as is the epicyclic frequency (Toomre 1964), while in a disk of gas and stars
 \be
  Q_{\rm gas + stars} = \frac{Q_{\rm gas} {\Sigma_g}/{\Sigma_0}} {{\rm max} \left[ \frac{\Sigma_g}{\Sigma_0} \frac{2 q  }{1+q^2}+ \frac{\Sigma_*}{\Sigma_0} \frac{2 q R}{1+R^2 q^2} \right]},
\label{eq:Qtot}
  \ee
where $R=\sigma_*/\sigma_{\rm 1D}$ is the ratio of stellar  and gas velocity dispersions, $\Sigma_* =  \Sigma_0-\Sigma_g$ is the stellar surface density, and the
denominator is evaluated at its maximum as a function of wavenumber $k$, with  $q \equiv k \sigma_{\rm 1D}/\kappa.$
(Jog \& Solomon 1984; Rafikov 2001).    In nearby disk galaxies $\Sigma_*/\Sigma_0 \geq 1$ and $R \geq 1,$ or 
$\Sigma_*/\Sigma_0 \leq 1$ and $R \geq 10$ (Leroy \etal 2008; Romeo \& Wiegert 2011).  
Over this full range, the denominator above varies from $\approx 0.7-1$, becoming very close to 1 if $R \approx 1$ or $\Sigma_* \gg \Sigma_0$, as is often the case in starbursts.
Thus I approximate $Q_{\rm gas + stars}$ as $Q_{\rm gas} \Sigma_g/\Sigma_0$ and set $Q_{\rm gas+stars} \approx 1,$
to give:
\be
\sigma_{1D} \approx \frac{f^{1/2} \pi G \Sigma_{0}}{\kappa}.
\label{eq:Q}
\ee

 The motivation for this requirement is both  theoretical, because $Q_{\rm gas + stars} \lesssim 1$ leads to 
 stirring of turbulence by star formation and spiral density waves (\eg Binney \& Tremaine 2008) and observational,  because $Q_{\rm gas + stars}  \approx 1-2$  is measured over a wide range of disk galaxies (Leroy \etal 2008).   From eqs.\  (\ref{eq:he}), (\ref{eq:pturb}), and (\ref{eq:Q}), this distribution will have a vertical scale length $H \approx \sigma_{\rm 1D} /(4 f^{1/2} \kappa).$
For spherical systems, an analogous requirement can be derived directly from hydrostatic equilibrium as
$\rho_g \sigma_{1D}^2/f  \approx 4 \pi G \Sigma_0 \Sigma_g $ implies $\sigma_{1D} \approx { f^{1/2} \pi G \Sigma_{0} t_{\rm ff}}$ and $H \approx \sigma_{1D}  t_{\rm ff}/(4 f^{1/2}),$
where $H$ and $t_{\rm ff} \equiv \sqrt{4/\pi G \rho_0}$ are the radius and the free-fall time.

Together these three properties lead to a further constraint. Turbulence is continuously cascading to smaller scales and thermalizing on a eddy turnover time, $L_t/\sigma_{\rm 3D},$ where $\sigma_{\rm 3D}=3^{1/2} \sigma_{\rm 1D}$ and $L_t$ and is the length scale of the largest eddies.  This means that turbulence must be continuously driven, and furthermore that thermal energy must be continuously removed from the system.   In a galaxy, this removal occurs through radiative cooling, which
can be described by a cooling function $\Lambda$ that depends on both gas temperature $T$ and metallicity $Z.$ 
This gives $\Lambda(T,Z) \left( {\rho_g}/{\mu m_p} \right)^2  \approx {\rho_g \sigma_{\rm 3D}^3}{L_T}^{-1} \ge {\rho_g \sigma_{\rm 3D}^3}{H}^{-1},$
where  $\mu m_p$ is the average particle mass and
the inequality arises because  the largest turbulent eddies must be smaller than the scale height.
Rearranging this  in terms of the $\Sigma_g$ gives:
\be
\Lambda(T,Z) \Sigma_g \ge 3^{3/2} (\mu m_p)^2 \sigma_{\rm 1D}^3,
\label{eq:LambdaT}
\ee 
a constraint that is completely independent of how turbulence is driven.

Furthermore, the cooling function cannot grow without bound, but rather it is divided into two regimes. Below $10^{5.5}$K, $\Lambda$ increases strongly with temperature,
reaching a maximum of $\Lambda^{\rm max} \approx 10^{-21}$ erg cm$^3$ s$^{-1}$ for material with $Z \approx Z_\odot.$  In this range, gas cooling is relatively stable, and a density perturbation in pressure equilibrium will cool at a similar rate as its surroundings.   On the other hand, above $10^{5.5}$K, a density perturbation will cool much quicker than the surrounding gas.  This regime is  unstable and cannot be maintained indefinitely.  Rather, turbulence will amplify density contrasts, leading to a distribution of cold clumps imbedded within a hot diffuse medium (Mathews \& Bregman 1978; Fall \& Rees 1985).

In this case, within the hot regions, cooling becomes increasingly less efficient because both density drops and $\Lambda(T,Z)$  decreases with increasing temperature. The result is that the temperature grows to several times the post-shock temperature for a $v=\sigma_{\rm 3D}$ shock, increasing the sound speed to well above the escape velocity and leading to a global outflow.   This turbulent runaway  has been simulated in Scannapieco, Gray, \& Pan (2012; hereafter SGP12) and it is fundamentally different from the picture of  a hot superbubble that bursts out from within a cold medium (\eg Mac Low \& Ferrara 1999; Silich \& Tenorio-Tagle 2001).   Instead, hot gas is continuously generated  over many dynamical times and (although the numerical diffusion in SGP12 was too high to demonstrate this conclusively) the 
 hot and cold gas phases are likely to exist cospatially at all times, as different aspects of the same multiphase, unstable turbulent medium.
The  $10^{5.5}$K dividing line at which turbulent runaway drives an outflow corresponds to $\sigma_{1D} \approx 35$ km/s (SGP12).  From eq.\  (\ref{eq:Q}), this  immediately gives a critical surface density  above which outflows will be generated of
\be
\Sigma_{0}^{\rm outflow}
\approx 
\approx 
\cases{300   \kappa_{10}  \,\,\, f^{-1/2} {\rm M_\odot pc}^{-2} & Disk \cr 
	    300   t_{\rm ff,10}^{-1}   f^{-1/2} {\rm M_\odot pc}^{-2} &  Spherical,}
\ee
where $\kappa_{10} = 10 \, {\rm Myrs} \times \kappa$ and $t_{\rm ff,10}= t_{\rm ff} /10$ Myrs.

For disk galaxies, Kennicutt (1998) found that $\dot \Sigma_\star  = 2.5 \pm 0.7 \times 10^{-4}  M_\odot \,  {\rm yr}^{-1}  {\rm kpc}^{-2} \, \left(\Sigma_{\rm g}/ {1 M_\odot \,  {\rm pc}^{-2}} \right)^{1.4 \pm 0.15} $  and  $\dot \Sigma_\star = 0.017 \Sigma_{\rm g} \Omega,$ where $\Omega$ is the circular velocity.
For a flat rotation curve one can use these fits to obtain $\kappa_{10} =  1.3 \left(\Sigma_{\rm g}/ {100 M_\odot \,  {\rm pc}^{-2}} \right)^{2/5},$ and  a fit to the data of Leroy \etal (2008), Table 7, gives $\Sigma_g/\Sigma_0  = \frac{1}{11 \pm 5} \left(\Sigma_{0}/ {100 M_\odot \,  {\rm pc}^{-2}} \right)^{-1/4}$ near $\Sigma_{0}^{\rm outflow},$ such that
$\Sigma_{0}^{\rm outflow}  = (200 \pm 50) f^{-5/7} {\rm M_\odot pc}^{-2}.$

\section{Implications}

\subsection{Implications for Starbursting Galaxies}

The generation of galaxy outflows by unstable turbulent support immediately leads to several important implications.  
Assuming that thermal pressure, turbulence, and magnetic fields are in rough equipartition,  $f \approx 1/3,$ 
we have $\Sigma_{g} \gtrsim  55 \pm 35 M_\odot \,  {\rm pc}^{-2}$ and
$\dot \Sigma_\star \gtrsim   0.08 \pm 0.06 M_\odot \, {\rm yr}^{-1} \, {\rm kpc}^{-2},$ which  is in excellent agreement with
$\dot \Sigma_\star^{\rm crit} \approx 0.1 M_\odot \, {\rm yr}^{-1} \, {\rm kpc}^{-2}.$
This limit corresponds  to a minimum pressure in an outflowing starburst of 
$n T \approx 7 \times 10^{6}  {\rm K} \, {\rm cm^{-3}},$ and 
unlike models that compare the energy input to the gravitational binding energy (e.g. Dekel \& Silk 1986),  it does not depend on galaxy mass, which explains the presence of outflows in massive  Lyman-break galaxies (Shapley et al. 2003), luminous and ultraluminous infrared galaxies (Martin 2005), and blue-cloud star-forming galaxies (Weiner \etal 2009). Only two previous papers  have attempted to explain this limit: Shu, Mo, \& Mao (2005) who associate it with a velocity of 160 km/s below which they claim winds are undetectable (see however Martin 2005), and Murray, M\' enard, \& Thompson (2011), who associate it with the value at which giant molecular clouds disrupt at velocities above the galaxy escape velocity. This, unlike our model, predicts a strong scaling of $\dot \Sigma_\star^{\rm crit}$ with galaxy mass.  

Secondly, our model predicts that the turbulent velocity scales roughly with $\Sigma_0,$ and should be at least 35 km/s in all starbursts.    
Recently, Genzel \etal (2011) compiled a large sample of the higher-redshift data and found that all galaxies with $\dot \Sigma_\star \ge \dot \Sigma_\star^{\rm crit}$ 
do in fact have $\sigma_{\rm 1D,H\alpha} \geq 35$ km/s.  

Thirdly, in our model, the ISM of all outflowing galaxies  contains gas with a wide range of temperatures in rough pressure equilibrium, including gas above  a few million degrees, the post-shock temperature of gas colliding at a few times the mean critical  turbulent velocity (SGP12). Again, these features are well supported observationally.  For example, recent  observations of M82 uncover a medium in which $\approx 10^4$ K dense clumps are surrounded by  $\gtrsim 10^7$K gas. In both components, $n T \approx 2 \times 10^7$ K cm$^{-3}$ (Strickland \& Heckman 2007). Note that this value, which is similar to that seen in other starbursting galaxies (Heckman \etal 1990), is a few times greater than the minimum pressure for turbulent runaway, suggesting that  if SGP12 had included supernova material (\eg Hill \etal 2012) in addition to turbulence, even hotter gas may have added to the hot medium initially opened up by turbulent runaway (Strickland \& Heckman 2009).

\subsection{Implications for Galaxy Outflows}

For most galaxies,  this hot gas will move many scale heights per Myr, while a typical starburst continues for many Myrs (\eg Greggio \etal 1998; F{\"o}rster Schreiber \etal 2003).   Thus the hot gas distribution should be well approximated by an equilibrium configuration. Such a solution was first derived for a constant energy and mass input rate within a spherical distribution (or equivalently for a conical distribution with a fixed opening angle) by Chevalier and Clegg (1985),
providing model that is often applied for simple estimates.  

This can be significantly improved as described in Silich \etal (2011), by accounting for the profile of the turbulent gas distribution that adds energy and gas to this wind. In this case the equations of mass, momentum, and energy conservation are:
\be
\frac{1}{r^2} \frac{d}{dr} (\rho_h v_h r^2) = \dot q_{\rm m}, 
\ee
\be
\rho v_h \frac{dv_h}{dr} = - \frac{dP_h}{dr} - \dot q_{\rm m} v_h,
\label{eq:mom}
 \ee
\be
\frac{1}{r^2} \frac{d}{dr} \left[ \rho_h v_h r^2 \left( \frac{v_h^2}{2} + \frac{\gamma}{\gamma -1} \frac{P_h}{\rho_h} \right) \right] =  \dot q_{\rm e},
\ee
where $\rho_h$, $v_h$, $P_h,$ and $\gamma$ are the density, radial velocity,  and ratio of specific heats of the hot gas.
The mass and energy input per unit volume are  given by truncated exponential profiles:
\be
\dot q_{\rm m} =  
\cases{ \dot q_{\rm m,0} \exp(-r/r_c)  & if r $\leq r_{\rm max}$ \cr
	        0, & if r $> r_{\rm max},$ \cr}
\ee
and $\dot q_{\rm e,0} =  \dot q_{\rm m,0} c_{0,h}^2/ (\gamma-1),$ where $c_{0,h} \gtrsim 500$ km/s is the sound speed of the hot gas at $r=0,$ and $r_c$, $r_{\rm max}$ are the scale radius and truncation radius, respectively.


\begin{figure*}
\epsscale{1.1}
\plotone{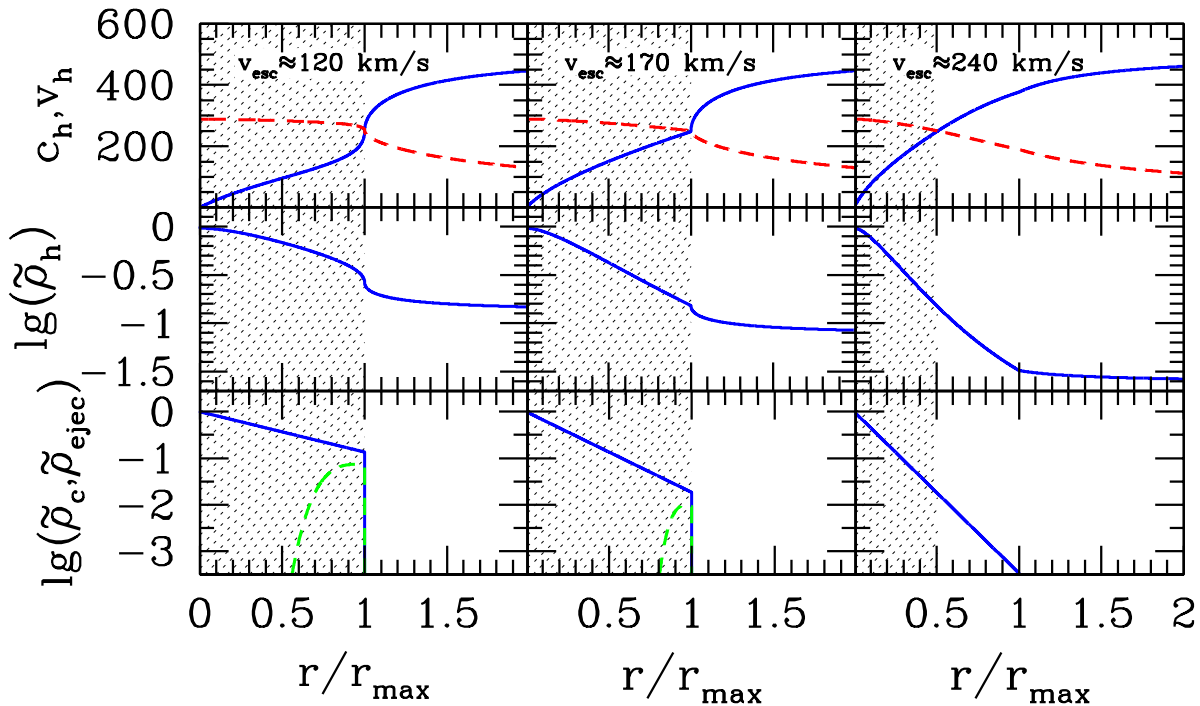}  
\caption{Equilibrium wind solutions with truncated exponential mass and energy input profiles.  From left to right the columns correspond to models  with truncation radii, $r_{\rm max} = 2 r_c$, $4 r_c$, and $8 r_c$, respectively,  labeled by their approximate escape velocities
$v_{\rm esc} \approx \sigma_{\rm 1D} \sqrt{2 r_{\rm max}/f r_c}, \approx$ 85 km/s $\sqrt{r_{\rm max}/r_c}$ where I assume $\sigma_{\rm 1D} \approx 35$ km/s and $f=1/3.$
{§\em Top row:} Radial velocity (solid) and sound speed (dashed) of the hot gas as a function of radius.   The shaded regions show the gas within the sonic point, which lies at the truncation radius if $r_{\rm max} \lesssim 4 r_c$ and at lies $\approx 4 r_c$ for larger $r_{\rm max}$ values.   {\em Center row:} Normalized hot gas density profiles.  {\em Bottom row:}  Normalized density of the cold gas reservoir (solid), and the cold gas expected to be ejected by turbulent motions (dashed).}
\vspace{0.12in}
\label{fig:evolution01}
\end{figure*}

These equations can be easily solved, yielding the models shown in Figure 1.
In the top row of this figure, I plot the hot gas velocity, $v_h,$ and sound speed $c_h$ as a function of radius for models with $r_{\rm max}$ = $2 r_c$ through $8 r_c$.  For small $r_{\rm max}$ values, the wind properties change dramatically at $r_{\rm max},$ while for large  $r_{\rm max}$ values, the  radial trend is more gradual.  In the $r_{\rm max}$ = $2 r_c$ model, for example, the sound speed within the driving region is roughly constant, but drops off quickly outside of the sonic point at $r=r_{\rm max}.$  In models with larger truncation radii, $c_h$ drops more significantly within $r_{\rm max}$ and the transition at $r_{\rm max}$ becomes less pronounced. Finally,  when $r_{\rm max} \gtrsim 4 r_c$ the sonic point  stays fixed at $\approx 4 r_c,$ rather than moving out to  $r_{\rm max}.$

The center row of this figure shows the normalized density profile of the hot gas $\tilde \rho_h(r) \equiv \rho_h(r)/\rho_h(0).$  Unlike the sound speed and velocity, $\tilde \rho_h(r)$ shows a dramatic change at  $r_{\rm max}$ for all values of $r_{\rm max}/r_c.$  In fact, the model with the largest truncation radius shows a sharp change in $\tilde \rho_h(r)$ at $r_{\rm max},$ but no feature at all at the sonic point.  Note that increasing the sound speed at $r=0$ would  have no effect on $\tilde \rho_h(r),$ and would raise both $c_s(r)$ and $v_h(r)$ by a fixed factor without changing their radial dependence.  

Finally, the bottom panel of Figure 1 estimates the cold gas ejection from these galaxies.   While the cold gas found in galaxy outflows is thought to be accelerated by the hot wind (\eg Martin 2005),  there is a serious problem with this picture.
Klein et al.\  (1994) studied the evolution of nonradiative cold clouds propagating at a velocity $v$ through a hot medium, showing that if $v$ is much greater than the cloud's internal sound speed, it will be shredded on a ``cloud crushing" timescale
\be
  t_{\rm cc} = \frac{\chi^{1/2} R_{\rm cloud}}{v},
\ee
where $\chi = \rho_{\rm cloud}/\rho_h$ is the density contrast between the media, and $R_{\rm cloud}$ is the cloud radius.  Subsequent 
studies showed that  magnetic fields   (\eg Mac Low et al.\ 1994) and  radiative cooling (Fragile et al.\ 2005) 
can only delay this disruption by 1-2 cloud crushing times.    On the other hand,  to accelerate the cloud to the hot wind velocity,
the impinging mass must be comparable to its total mass, which takes a time
\be
t_{\rm accel} = \frac{(4 \pi/3) R_{\rm cloud}^3 \rho_{\rm cloud}}{\pi R_{\rm cloud}^2 \rho_h  v}  
= \frac{ 4}{3} \chi^{1/2} t_{\rm cc}.
\label{eq:taccel}
\ee
Because they are in rough pressure equilibrium (Strickland \& Heckman 2007), the density contrast between the $\lesssim 10^4$K clouds and the hot gas is $\chi \approx T_{\rm h}/T_{\rm cold} \gtrsim 1000,$ such that $t_{\rm accel}$  exceeds 30 cloud crushing times.  Given the results discussed above, it would appear that cold clouds will never survive to reach velocities comparable to that of the hot wind.

There is, however, one likely caveat.  If cloud cooling is efficient {\em and the relative velocity exceeds the hot gas sound speed,}  a bow shock develops in front of the cloud. This protects the cloud from ablation, both by reducing heating within the cloud and shear around its sides.  When combined with cooling, these mitigating effects allow the cloud to remain intact for many cloud crushing times.    In fact,  the  acceleration of clouds  without strong disruption in this case has been simulated in both Cooper \etal (2009) and Kwak \etal (2011), although these runs were terminated before the cloud velocity approached $v_h.$  Nevertheless, we can infer that the acceleration of cold clouds is likely to be efficient if they enter the hot gas outside of the sonic point, and inefficient if they enter the flow where $v_h < c_h.$    

In Figure 1, we see that
only galaxies with escape velocities $\gtrsim 200$ km/s have a portion of their cold gas reservoir outside of the subsonic region.   For these, one would observe many cold gas clouds with velocities comparable to $v_h$ outside of the sonic point, but significant shredding of cold clouds  inside of the sonic point.  In fact, the archetypal high-mass starburst, M82, displays exactly these features.  Within 300 pc of the disk, X-ray observations indicate significant mass loading of the hot wind (Suchkov \etal 1994) and the cold gas motions are minimal (McKeith \etal 1995).  Outside 300 pc,  the cold gas motions quickly approach $v_h,$ and strong broad line SII emission is observed (Westmoquette \etal 2009), indicating a large population of cold clouds interacting with the hot outflow.   

Similarly, the fastest cold gas velocity as measured by NaI absorption approaches $v_h$  for  galaxies with circular velocities $\gtrsim 150$ km/s (Martin 2005), while the fastest cold gas velocity is $\lesssim $100 km/s in lower mass galaxies (Martin 2005).  These correspond to models with $r_{\rm max} < 4 r_c$, for which there is no cold reservoir beyond the sonic point.  However, these potential wells are small enough that  the stochastic turbulent motions themselves will accelerate a fraction of the gas to above the escape velocity.  The lower panel of Figure 1 shows the profile of cold gas expected to be ejected by turbulence, calculated simply as $\tilde \rho_{\rm eject}(r) =  \tilde \rho_c(r) \, 0.5 \, {\rm erfc}[v_{\rm esc}(r) / \sqrt{2 \sigma_{\rm 1D}^2}],$ where I crudely approximate $v_{\rm esc}(r)$ as  $v_{\rm esc}(0) \sqrt{1-r/r_{\rm max}}.$   For these galaxies, it is easy to understand the often perplexing observation that the hot gas is likely to be moving at $\gtrsim 500 $ km/s while the ejected cold clouds are moving $\lesssim 100 $ km/s (\eg Martin 2005).   These are young clouds ejected by  ISM turbulence, and they have not yet experienced significant acceleration by the wind.  If they had, they would have already been shredded.

\section{Extreme Surface Density Limit}

Finally, there is an extreme regime, in which $\Sigma_0$ is simply too high to be stable to the Toomre criterion with significant turbulent support, while at the same time balancing  energy input with cooling, even if $\Lambda = \Lambda^{\rm max}.$  In this case eqs.\  (\ref{eq:Q}) and (\ref{eq:LambdaT}) cannot be simultaneously satisfied for a single dynamical time, and there is no way for the gas to achieve even an unstable equilibrium.   For disk and spherical configurations this occurs when the surface density is above
\be
\Sigma_0^{\rm max}  \approx 
\cases{3 \times 10^5 \, \kappa_{10}^{3/2} \, (\Sigma_g/\Sigma_0)^{1/2} M_\odot \, {\rm pc^{-2}} & Disk \cr 
	    3\times 10^5 \, t_{\rm ff,10}^{-3/2} \, (\Sigma_g/\Sigma_0)^{1/2} M_\odot {\rm pc^{-2}} &  Spherical,}
\ee
where I have taken $f=1/3$.   

These values closely correspond to the maximum observed stellar surface density of $\Sigma_{\rm max} \approx 10^ 5 M_\odot \, {\rm pc}^{-2},$
which is consistent across globular clusters, nuclear star clusters, super star clusters, ultra-compact dwarfs, and the centers of
elliptical galaxies (Hopkins \etal 2010).   Above $\Sigma_0^{\rm max},$ gas can only be driven out  within a dynamical time or fall freely without achieving equilibrium.  In fact, Larson (2009) showed that if gas above this density were deposited onto a central supermassive black hole, its mass would match the observed relation to the stellar velocity dispersion.  Perhaps, like galaxy outflows, this is a consequence of the physics of turbulent support at high surface densities.

\acknowledgements

I would like to thank  Yan-Mei Chen, Timothy Heckman, Mark Krumholz, Sangeeta Malhotra, Crystal Martin, Liubin Pan, Christy Tremonti, and Mark Westmoquette  for discussions that greatly improved this manuscript.   I acknowledge the support of NASA theory grant NNX09AD106 and NSF grant AST11-03608.

\end{document}